\documentclass[fleqn,usenatbib]{mnras}

\usepackage{newtxtext,newtxmath}

\usepackage[T1]{fontenc}
\usepackage{ae,aecompl}

\usepackage{graphicx}	
\usepackage{amsmath}	
\usepackage{amssymb}	
\usepackage{verbatim}
\usepackage[export]{adjustbox}
\usepackage{times}

\def \prd {{\it Phys. Rev. D}}

\def \apj {{\it ApJ}}
\def \apjl {{\it ApJ Let.}}
\def \apjs {{\it ApJ Sup.}}

\def \aap {{\it A\&A}}

\def \mnras {{\it MNRAS}}
\def \nat {{\it Nature}}

\def\physrep{{\it Phys.~Rep.}}

\title[When galaxies align]{When galaxies align: intrinsic alignments of the progenitors of elliptical galaxies in the Horizon-AGN simulation}

\author[J. Bate et al.]{
\parbox{\textwidth}{
James Bate,$^{1}$
Nora Elisa Chisari,$^{2}$\thanks{E-mail: n.e.chisari@uu.nl}
Sandrine Codis,$^{3}$
Garreth Martin,$^{4,5,6}$\\
Yohan Dubois,$^{3}$
Julien Devriendt,$^{1}$
Christophe Pichon,$^{3,7}$
and Adrianne Slyz.$^{1}$}
\\
\vspace*{6pt}\\
\noindent
$^{1}$Department of Physics, University of Oxford, Keble Road, Oxford OX1 3RH, United Kingdom.\\
$^{2}$Institute for Theoretical Physics, Utrecht University, Princetonplein 5, 3584 CC Utrecht, The Netherlands.\\
$^{3}$Institut d'Astrophysique de Paris, CNRS \& Sorbonne Universit\'e, UMR 7095, 98 bis Boulevard Arago, 75014, Paris, France.\\
$^{4}$Centre for Astrophysics Research, School of Physics, Astronomy and Mathematics, U. of Hertfordshire, College Lane, Hatfield AL10 9AB, UK.\\
$^{5}$Steward Observatory, University of Arizona, 933 N. Cherry Ave, Tucson, AZ, USA.\\
$^{6}$Korea Astronomy and Space Science Institute, 776 Daedeokdae-ro, Yuseong-gu, Daejeon 34055, Korea.\\
$^{7}$ Korea Institute for Advanced Study, 85 Hoegiro, Dongdaemun-gu, Seoul, 02455, Republic of Korea.
}

\date{Accepted 2019 November 07. Received 2019 September 10; in original form 2019 January 22.}

\pubyear{2019}

\begin{document}
\label{firstpage}
\pagerange{\pageref{firstpage}--\pageref{lastpage}}
\maketitle

\begin{abstract}
Elliptical galaxies today appear aligned with the large-scale structure of the Universe, but it is still an open question when they acquire this alignment. Observational data is currently insufficient to provide constraints on the time evolution of intrinsic alignments, and hence existing models range from assuming that galaxies gain some primordial alignment at formation, to suggesting that they react instantaneously to tidal interactions with the large-scale structure. Using the cosmological hydrodynamical simulation Horizon-AGN, we measure the relative alignments between the major axes of galaxies and eigenvectors of the tidal field as a function of redshift. We focus on constraining the time evolution of the alignment of the main progenitors of massive $z=0$ elliptical galaxies, the main weak lensing contaminant at low redshift. We show that this population, which at $z=0$ has a stellar mass above $10^{10.4}$ M$_\odot$, transitions from having no alignment with the tidal field at $z=3$, to a significant alignment by $z=1$. From $z=0.5$ they preserve their alignment at an approximately constant level until $z=0$. We find a mass-dependence of the alignment signal of elliptical progenitors, whereby ellipticals that are less massive today ($10^{10.4}<M/{\rm M}_\odot<10^{10.7}$) do not become aligned till later redshifts ($z<2$), compared to more massive counterparts. We also present an extended study of progenitor alignments in the parameter space of stellar mass and galaxy dynamics, the impact of shape definition and tidal field smoothing.
\end{abstract}




\section{Introduction}

Elliptical galaxies are known to align with the large-scale structure of the Universe. Observational evidence of this phenomenon was first identified in the pioneering work of \citet{Binggeli82} in an analysis of Brightest Cluster Galaxies in the Abell sample of clusters. Since then, many works have shown a tendency for luminous red galaxies to align radially towards over-densities in the matter field \citep{Hirata04b,Mandelbaum06,Hirata07,Joachimi11,Singh14,2017MNRAS.468.4502V,Johnston18}, with more luminous or massive galaxies displaying a stronger signal. 
There is also evidence of them having a preferential orientation with respect to filaments in the cosmic structure, aligning their minor axes in the direction perpendicular to the filament \citep{Tempel13} or their major axes parallel to them \citep{Chen19}. Their intrinsic angular momenta (``spins'') have also been found to align with the cosmic shear field: parallel to the axis of greatest compression and perpendicular to the axis of slowest compression \citep{Pahwa16}. These alignments are induced by the large-scale tides produced by the cosmic web (clusters, filaments, walls and voids). The large-scale tidal field indeed generates on the one hand tidal stretching \citep{Catelan01} and on the other hand, tidal torquing (see \citealt{Schaefer09} for a review) spinning up haloes and galaxies in a cosmic-web dependent way \citep{codis15b}.
With the emergence of weak gravitational lensing surveys, and their application to precision cosmology \citep{Huff14,Troxel17,vanUitert18,Joudaki18}, intrinsic correlations between galaxy shapes (``intrinsic alignments'') were identified as a potential contaminant to the lensing signal \citep{Brown02,Hirata04}. (For an overview of intrinsic alignments, see \citealt{Joachimi15,Troxel15}.)

With the goal of mitigating contamination to gravitational lensing observables, models of intrinsic alignment correlations were proposed which connect the projected shape of galaxies to the tidal field of the large-scale structure \citep{Catelan01,Mackey02}. These models have been successful in reproducing current observations \citep{Joachimi11,Singh14}, but the redshift evolution of the intrinsic shape correlations remains poorly constrained. Assumptions in the models range from galaxies reacting instantaneously to the tidal field to alignments being set up at some ``primordial'' redshift when the galaxy formed. A prior on redshift evolution of intrinsic alignments would greatly improve the performance of mitigation strategies \citep{Kirk12,Krause15}.

A viable strategy for obtaining a prior on redshift evolution of alignment models is to use numerical simulations. In particular, recent cosmological hydrodynamical simulations have been successful in predicting the alignment trends of low redshift ellipticals \citep{Ten++14,Tenneti15a,Tenneti15b,Velliscig15,Velliscig15b,Chisari15,Chisari16,Hilbert17}. 

In this work, we use a cosmological simulation, Horizon-AGN \citep{Dubois14} to follow the evolution of elliptical galaxy alignment with the tidal field. Knowing that the simulation reproduces observed trends of the alignments of massive ellipticals, we ask the question of when this population gained its alignment with the tidal field. To answer it, we construct a merger tree that allows us to track the main progenitors of redshift $z=0$ ellipticals as a function of time up to $z=3$, covering the range of interest of future lensing surveys such as the Large Synoptic Survey Telescope (LSST\footnote{\url{http://lsst.org}}), {\it WFIRST}\footnote{\url{https://wfirst.gsfc.nasa.gov}} or {\it Euclid}\footnote{\url{http://sci.esa.int/euclid/}}. We choose this method form a purely theoretical perspective, as of course such an excercise would be impossible to perform in real data. Our results suggest that elliptical galaxies with a stellar mass above $10^{10.4}$ M$_\odot$ at low redshift gain their alignment by $z\simeq 1$ and preserve it thereafter. For a reader interested in direct comparisons between Horizon-AGN alignments in projection to observational constraints, we refer them to \citet{Chisari15,Chisari16}.

This work is structured as follows. Section \ref{sec:sim} describes the Horizon-AGN cosmological hydrodynamical simulation, in particular, features that are relevant to this work. Further details on Horizon-AGN can be obtained from \citet{Dubois14}. Section \ref{sec:methods} describes our methods for quantifying intrinsic alignments and our choice of galaxy sample. We present our results in Section \ref{sec:results}. These are compared to previous work in Section \ref{sec:discuss} and we conclude in Section \ref{sec:conclusions}.

\section{Horizon-AGN simulation}
\label{sec:sim} 

The Horizon-AGN simulation is a cosmological hydrodynamical simulation performed with the adaptive-mesh-refinement code {\sc Ramses} \citep{teyssier02}. The dimensions of the simulation box are $L=100 \, h^{-1}\rm\,Mpc$ on each side. The simulation follows the evolution of galaxies in the large-scale structure, modelling star formation, feedback from supernovae and Active Galactic Nuclei (AGN), ultraviolet background heating, gas cooling and stellar winds according to state-of-the-art recipes \citep{ greggio&renzini83,sutherland&dopita93,haardt&madau96,leithereretal99, leithereretal10,rasera&teyssier06,dubois&teyssier08winds,booth&schaye09,duboisetal12agnmodel}. 

The cosmological model adopted is a standard $\Lambda$CDM cosmology with parameters set by the {\it WMAP}7 results~\citep{komatsuetal11}, i.e.: a total matter density $\Omega_{\rm m}=0.272$, dark energy density $\Omega_\Lambda=0.728$, amplitude of the matter power spectrum $\sigma_8=0.81$, baryon density $\Omega_{\rm b}=0.045$, Hubble constant $H_0=70.4 \, \rm km\,s^{-1}\,Mpc^{-1}$, and $n_s=0.967$. The simulation follows the evolution of $1024^3$ dark matter (DM) particles, with a resulting mass resolution of $M_{\rm DM, res}=8\times 10^7 \, \rm M_\odot$. The adaptive mesh allows the simulation to reach a $\Delta x=1\, \rm kpc$ resolution in the densest regions of the box, and uses an approximate stellar mass resolution of $M_{*, \rm res}=\rho_0 \Delta x^3\simeq 2\times 10^6 \, \rm M_\odot$. Further details on the Horizon-AGN simulation can be found in \citet{Dubois14}. 

In the next sub-sections, we focus on describing the extraction of the simulated quantities of particular relevance to this work. 

\subsection{Galaxy shapes}
\label{sec:shapes}

Galaxies are identified in each redshift snapshot of the simulation using the {\sc AdaptaHOP} finder~\citep{aubertetal04}. This algorithm relies on the distribution of stellar particles to estimate the local density around each particle. Over-densities that exceed a local threshold of $178$ times the average total matter density and with more than 300 stellar particles are identified as galaxies with reliable estimates of their shapes \citep{Chisari15c}. 

Galaxy shapes are modelled as ellipsoids. The axes of the ellipsoids point in the directions of the eigenvectors of the inertia tensor, which is defined as the following sum over $n$ stellar particles,
\begin{equation}
I_{ij} = \frac{1}{M}\sum_n m_{(n)}x_i^{(n)}x_j^{(n)},
\label{eq:sit}
\end{equation}
where $i,j = {1,2,3}$ correspond to the axes of the simulation box, $m$ is the mass of the stellar particle, $x$ is the distance of each particle to the centre of mass of the galaxy, and $M$ is the total stellar mass of the galaxy. 
This tensor is diagonalized, and the eigenvalues (labelled $c,b,a$ from the smallest to the largest) correspond to the lengths of the minor, intermediate, and major axis, respectively.	We use the axis ratios $c/a$ and $b/a$ as a proxy for the ellipticity of the galaxies.

\subsection{Tidal Field Extraction}
\label{sec:tidalfield}

In this work, we focus on the alignment of simulated galaxies with the smoothed tidal field at their position throughout their cosmic evolution. The three-dimensional (traceless) tidal tensor is defined as
\begin{equation}
  T_{ij} = \partial_{ij} \Phi - \frac{1}{3} \Delta \Phi \,{\delta}_{ij}\,,
\end{equation}
where $\Phi$ is the gravitational potential, $\Delta\Phi$ is the Laplacian of the gravitational potential, and $\delta_{ij}$ the Kronecker delta function. 
Its minor, intermediate and major eigenvectors are labelled
$\mathbf{v}_1$, $\mathbf{v}_2$ and $\mathbf{v}_3$ and correspond, respectively, to the ordered eigenvalues denoted 
$\lambda_{1}\le\lambda_{2}\le\lambda_{3}$.

In practice, we compute the tidal tensor smoothed on scale $R_{s}$, $T_{ij}=\partial_{ij} \Phi_{R_{s}}-\Delta \Phi_{R_{s}}\,{\delta}_{ij}/3$, via a Fast Fourier Transform of the total density field that includes contributions from dark matter, stars, gas and black holes. This density field is sampled on a $512^{3}$ grid (corresponding to a comoving scale of $200$ kpc/$h$) and convolved with various Gaussian filters of comoving scale $R_{s}= 0.4$, $0.8$, $1.6$ and $3.2$ Mpc/$h$, allowing us to then estimate
\begin{equation}
\partial_{ij} \Phi_{R_{s}}(\mathbf{x})=\frac{3 H_0^2 \Omega_0}{2 a}\int \mathrm{d}^{3}\mathbf{k}\;\delta(\mathbf{k}) \frac{k_{i}k_{j}}{k^{2}}W_{G}(k R_{s})\exp\left({i \,\mathbf{k}\!\cdot\!\mathbf{x}}\right)\,, 
\end{equation}
where $\delta(\mathbf{k})$ is the Fourier transform of the sampled density field and $W_{G}$ a Gaussian filter. ($\Delta\Phi$ is similarly obtained from $\partial_{ij} \Phi_{R_{s}}$.) This procedure is applied at the following redshifts: $z=\{0.06,0.12,0.2,0.31,0.42,0.5,0.64,1,2,3\}$.
We thus use two sets of simulation outputs: a coarsely sampled set at $z=\{0.06,1,2,3\}$ that can give us insights into the broad redshift evolution of the alignment signal, and a more refined set at low redshift, with $6$ snapshots spanning the range $0<z<0.7$ where we expect a steep increase in the fraction of ellipticals according to our previous work \citep{Chisari16,Dubois16,Martin2018_progenitors}. 

Note that the tidal field  eigenvectors and their eigenvalues, are closely connected to the classification of the cosmic web into filaments, walls and knots of the large-scale structure \citep{zeldovich70,Bond96,Hahn07,Libeskind17}. For the purpose of this work, it suffices to recall that in filaments, the spine of the filament follows the direction of ${\bf v}_1$. Walls in the cosmic web have their planes determined by ${\bf v}_1$ and ${\bf v}_2$, and they are perpendicular to ${\bf v}_3$. In previous work, \citet{Codis18} studied the evolution of galaxy orientation with these elements of the cosmic web in the Horizon-AGN simulation. We connect our results to that work in Section \ref{sec:discuss}.

\subsection{Merger Tree Extraction}
\label{sec:mergertree}

Using the catalogue of galaxies identified by \textsc{AdaptaHOP}
at each snapshot, merger trees are extracted using the method
described by \citet{tweedetal09}. Merger trees are produced for each galaxy at the base snapshot ($z=0.06$), following their merger histories between the base snapshot and $z=3$. This allows us to track back in time the main (most massive) progenitor of each galaxy at evenly spaced time-steps of $\sim130~$Myr.
The choice of this specific timescale is driven by the fact that one would not expect the dynamical friction timescale of mergers of comparable mass to be $<200$ Myr \citep{Boylan08}. Hence, a 130 Myr timescale should be enough to properly track the progenitors of any given galaxy. As we further detail below, for our purposes it is unnecessary to track alignments on such a fine timescale and we focus only on selected redshift snapshots of the simulation among those for which progenitors are available.

For this work, we are particularly interested in the main progenitors of elliptical galaxies at $z=0.06$. Notice progenitors can be either elliptical or disc-like, as we do not impose restrictions on their stellar dynamics other than at $z=0.06$.

\section{Methods}
\label{sec:methods}

\subsection{Measuring alignments}
\label{sec:Measuring Alignments}

In this work, we investigate local alignments between galaxies and the tidal field in the Horizon-AGN simulation. To quantify this alignment, we use the angle between the eigenvectors of the tidal tensor, and the major axis of the galaxies because for elliptical galaxies, which are the subject of this work, the projected major axis is typically used in searching for observational alignment signatures\footnote{For discs, the minor axis is often better defined than the major axis \citep{Chisari15c}, but we opt to use the latter due to our focus on elliptical galaxies in this work.} \citep[e.g.][]{Binggeli82,Niederste10}.  Galaxies are ascribed the tidal tensor eigenvectors corresponding to their location in the grid constructed to obtain the tidal field (see Section \ref{sec:tidalfield}). The angle of each of the tidal field eigenvectors with the major axis of the galaxies is labelled $\theta_1^{\rm major}$, $\theta_2^{\rm major}$ and $\theta_3^{\rm major}$, respectively. 

To measure the alignment of the galaxies, we compare the distribution of $\theta_1^{\rm major}$, $\theta_2^{\rm major}$ and $\theta_3^{\rm major}$ angles to the random expectation. The probability density function corresponding to random alignment is $f(\theta_i^{\rm major}) = \sin(\theta_i^{\rm major})$, where we normalise over $\theta = [0,\pi/2]$. If we make a change of variable $u = \cos(\theta_i^{\rm major})$, a random alignment trend then corresponds to $f(u) = 1$. In the figures that follow, we show the probability density distribution resulting from the random contribution plus any excess, i.e. $f(u)=1+\xi(u)$. Similarly, if the alignments of the galaxies are random, we expect an average angle of $\langle\theta\rangle = \left(\int_{0}^{2\pi} d\phi \int_{0}^{\pi/2} \theta \sin{\theta} d\theta\right)/\left(\int_{0}^{2\pi} d\phi \int_{0}^{\pi/2} \sin{\theta} d\theta\right) = 1$. This corresponds to an angle of $\sim 57\deg$. The median of such distribution corresponds to $60\deg$.

The error bars for each of the curves presented are calculated as the Poisson standard error. To test the accuracy of these error bars we also performed an alternative estimation using eight same-volume sub-boxes in the simulation, and calculated the standard deviation of the mean for each bin. The error bars obtained from this method were similar to the Poisson errors, and hence the latter are adopted for all curves.

We also experimented with altering the size of the smoothing filter applied to the tidal field as discussed in Section \ref{sec:tidalfield}. We tested $R_s=$0.4, 0.8, 1.6 and 3.2 Mpc$/h$ Gaussian smoothing kernels, and found that this made no significant difference to the results (see Appendix \ref{app:smooth}). Hence, we will only show results for the $R_s=$0.4 Mpc$/h$ Gaussian smoothing in what follows. While our results are robust to these choices of smoothing scales, adopting scales larger than 3.2 Mpc$/h$ is expected to decrease the amplitude of the alignment strength, in accordance with linear theory predictions \citep{Catelan01}.

\subsection{Selection of elliptical galaxies and their main progenitors}

Table \ref{tab:galnum} shows the number of galaxies with reliable shapes identified in Horizon-AGN as a function of redshift for the outputs of interest in this work. This population, which includes disc-like galaxies at all redshifts, provides a reference alignment measurement to which we can compare the alignment of other galaxy samples.

\begin{table}
\caption{Number of galaxies in Horizon-AGN that pass our selection cut on number of stellar particles (first column) and restricting to the main progenitors of today's high mass ($M/{\rm M}_\odot>10^{10.4}$, second column) elliptical at different redshifts. The first row of the table at $z=0.06$ indicates the number of massive ellipticals for which their main progenitors are sought at higher redshifts in the subsequent rows.}
\begin{center}
\begin{tabular}{ |c|c|c| }
\hline
$z$ & All galaxies & High mass elliptical progenitors\\
\hline 
0.06 & 84499 & 4217\\  
0.12 & 85723 & 4105 \\ 
0.2 & 87244 & 4135\\ 
0.31 & 88632 & 4142\\  
0.42 & 90114 &4154\\  
0.5 & 91070 &4163\\  
0.64 & 91861 & 4179\\  
1 & 90456 & 4165\\ 
2 & 70665 & 4027\\ 
3 &  41168 & 3478\\  
\hline
\end{tabular}\label{tab:galnum}
\end{center}
\end{table}

From this population, we select elliptical galaxies at $z = 0.06$ by using the ratio $V_{\theta}/\sigma$ as a proxy for the dynamics of the galaxies. This ratio is defined in the following way: first we define a cylindrical coordinate system, with the $z$-axis parallel to the total angular momentum of the stars in the galaxy. This allows us to decompose the velocity of each star into $v_r$, $v_{\theta}$ and $v_z$ components. We then define $V_{\theta} = \overline{v_{\theta}}$, where the over-line refers to the average value. We also decompose the velocity dispersion into cylindrical coordinates $\sigma_r$, $\sigma_{\theta}$ and $\sigma_z$, with the total dispersion satisfying $\sigma^2 = (\sigma_r^2 + \sigma_{\theta}^2 + \sigma_z^2)/3$. Low values of $V_{\theta}/\sigma$ indicate that a galaxy is more elliptical, whereas high values indicate it is more disc-like. We define ellipticals at $z = 0.06$ as those galaxies with $V_{\theta}/\sigma < 0.6$. 

We prefer to rely on a purely dynamical criterion for our selection of ellipticals for several reasons. First, for consistency with our previous alignment studies \citep{Chisari15,Chisari16,Chisari2017,Codis18}. Second, because the alignment mechanism is hypothesized to be different for galaxies depending on their internal dynamics, and we want to make as reliable a distinction as possible between discs and ellipticals. Other proxies for morphology, such as colour, could contaminate our sample. Finally, \citet{Dubois16} found that implementing this criterion results in a good match between the fraction of ellipticals in Horizon-AGN and in observations (see their Figure 4 and the comparison to \citealt{Conselice06}). 

\begin{figure}
\includegraphics[width=\columnwidth,trim=0 0 20bp 25bp,clip]{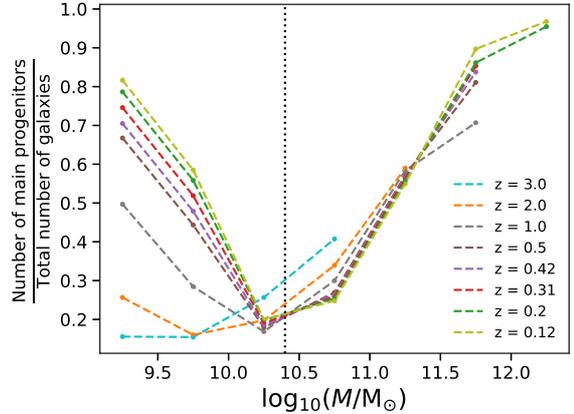}
\caption{Ratio of the number of $z=0.06$ elliptical progenitors as a function of mass to the full population of galaxies at different redshifts in the Horizon-AGN simulation. Both numerator and denominator are taken at the same redshift, indicated in the legend. A ratio of $1$ in this plot indicates that all galaxies in Horizon-AGN at a given mass bin are main progenitors of today's ellipticals. The $z=0.06$ curve displays two peaks at low and high mass, suggesting that most low mass and high mass galaxies at this redshift are ellipticals. On the contrary, at high redshift (e.g., $z=3$), only massive galaxies tend to be labelled as main progenitors of today's ellipticals. The black dotted vertical line indicates the mass threshold adopted to remove low mass ellipticals at $z=0.06$.}
\label{fig:masshist}
\end{figure}

Figure \ref{fig:masshist} shows the ratio of the mass distribution of elliptical progenitors to the full population of galaxies at any given redshift. In other words, this figure gives us an answer to the question of how special elliptical progenitors are as a function of redshift.
The $z=0.06$ corresponds to all massive ellipticals at this redshift, rather than their progenitors. It can be clearly seen that ellipticals are abundant at both low and high masses, and a subdominant population at intermediate masses. The low mass excess is in disagreement with observations, potentially as a consequence of lack of resolution below $1$ kpc \citep{Kaviraj17}. For this reason, and considering that observational works find a strong mass dependence of the alignment signal \citep{vanUitert18}, with more massive galaxies displaying larger alignment amplitudes, we focus on tracing the alignment history of $\log_{10}(M/{\rm M}_{\odot}) > 10.4$ ellipticals only (dotted vertical line). The threshold of $\log_{10}(M/{\rm M}_{\odot}) > 10.4$ was chosen in agreement with the findings of \citet{Dubois16} and \citet{ Martin2018_progenitors}, who showed that the fraction of ellipticals above this mass threshold from the Horizon-AGN simulation matches existing observations.
We also discard from our $z = 0.06$ sample a small number of galaxies identified as sub-structure of larger galaxies by {\sc AdaptaHOP} ($\lesssim 10\%$). This improves the completeness of the main progenitor samples, though the final results on alignment trends are not impacted. For a discussion of the impact of sub-structure on estimated alignments in Horizon-AGN, see \citet{Chisari16}.

We identify the main progenitors at the different redshifts of interest using the merger tree described in Section \ref{sec:mergertree}. The resulting number of galaxies at each redshift for this selection criteria is quoted in the third column of Table \ref{tab:galnum}. The redshift evolution evidenced in Figure \ref{fig:masshist} suggests that by $z=2$ there are few low mass main progenitors in our sample. This is expected, as low mass galaxies have typically formed more recently than high mass ellipticals. At $z=3$, most progenitors are high mass.

\begin{figure}
\includegraphics[width=\columnwidth]{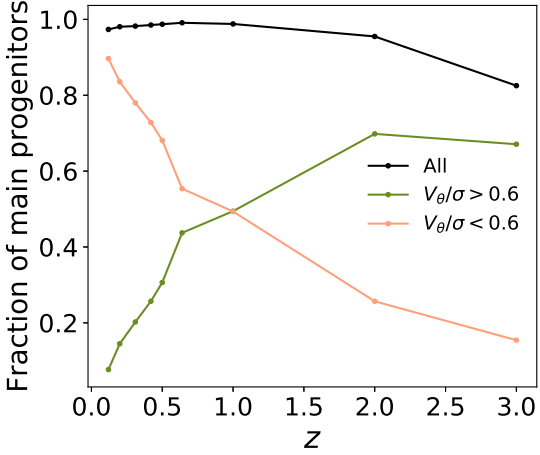}
\caption{The fraction of main progenitors of massive ellipticals at $z=0.06$ identified by the merger tree at each redshift. The black line shows the completeness of the merger tree, and the green and orange lines show the fraction of discs ($V_\theta/\sigma>0.6$) and ellipticals ($V_\theta/\sigma<0.6$) identified, respectively. Due to no restriction being placed on the dynamical properties of the main progenitor galaxy, the fraction of discs increases towards high redshift. Thus, at $z=3$, the main progenitors of today's massive ellipticals tend to be disc-like. Note that there is also no restriction on the mass of the progenitors, though they have to be the main (most massive) one at each redshift. 
}
\label{fig:completeness}
\end{figure}

Figure \ref{fig:completeness} shows the completeness of the sample of main progenitors at the different redshifts considered in this work. This is defined as the number of main progenitors identified at a given redshift divided by the original number of high-mass ellipticals in the parent sample at $z=0.06$. The resulting completeness remains above $95.9\%$ up to $z=2$ and only drops to $84\%$ for $z=3$. Because we place no restrictions on the dynamics of a progenitor, this sample is comprised both of discs and elliptical galaxies at $z>0.06$. As expected, the fraction of ellipticals in the main progenitor sample decreases towards high redshift in favour of an increased fraction of discs. The non-monotonic trend observed in the fraction of progenitors between $0<z<0.5$ (black solid curve in Figure \ref{fig:completeness}) is a consequence of the removal of sub-structures. Though the progenitor connection in the merger tree is never lost, sub-structures can have a small impact on completeness levels.   


To investigate the mass-dependence of progenitor alignment with the tidal field, we make a further split of the population of $z=0.06$ ellipticals into three mass bins with approximately equal number of galaxies. The boundaries of these mass bins are thus defined as $10.4 < \log_{10}(M/{\rm M}_{\odot}) <10.7$,
$10.7 < \log_{10}(M/{\rm M}_{\odot}) < 11.06$, and
$\log_{10}(M/{\rm M}_{\odot}) > 11.06$.

\section{Results}
\label{sec:results}

In this section, we examine the strength of the alignments of low redshift elliptical galaxies and their main progenitors with the tidal field as a function of redshift. For brevity, we only show alignment probability distributions for $\theta_1^{\rm major}$, the angle between the ${\bf v}_1$ eigenvector corresponding to the smallest eigenvalue, and the major axis of the galaxy, as defined in Section \ref{sec:Measuring Alignments}. 
In general, we find that the alignment signals of the major axes of galaxies with ${\bf v}_1$ and ${\bf v}_3$ are inverted. We interpret this as a consequence of the relative orientation of these eigenvectors with the cosmic web. ${\bf v}_1$ points parallel to filaments and walls, while ${\bf v}_3$ is perpendicular to these structures. Notice that we also find a weaker alignment signal for the major axes of our sample of galaxies with ${\bf v}_2$ than with the other eigenvectors of the tidal field. If alignments were only determined by the direction of a filament, we would expect them to be similar with respect to ${\bf v}_2$ and ${\bf v}_3$, as both of these eigenvectors are perpendicular to them. The fact that the alignment with respect to these two vectors is different suggests that physical processes {\it inside} walls play a significant role in determining the alignment of galaxies with the tidal field. Assuming that the spin and major axes of progenitors are typically perpendicular to each other (as confirmed in Appendix \ref{app:ms}), these results are in agreement with the findings of \citet{Codis18}. In that work, indeed it was found that galaxies have a significant alignment with respect to walls, with their spins either perpendicular to the normal to the wall (at low mass) or parallel to it (at high mass).

In what follows, we investigate the redshift evolution of these trends for the full sample of galaxies with reliable shapes in Horizon-AGN and for the sample of main progenitors of low redshift massive ellipticals as defined in the previous section. When investigating the redshift evolution of the mean alignment angle of the sample with redshift, we quote results for all alignment angles: $\theta_1^{\rm major}$, $\theta_2^{\rm major}$ and $\theta_3^{\rm major}$.

\subsection{Alignments of the full galaxy sample}

\begin{figure}
\centering
\includegraphics[width=1.2\columnwidth,trim=30 0 0 25bp,clip]{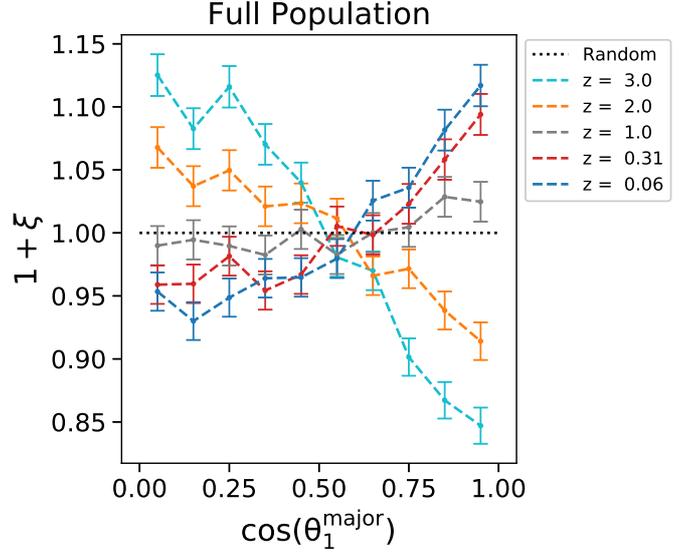}
\caption{Probability density distribution of $\cos(\theta_1^{\rm major})$, the cosine of the angle between the major axis of a galaxy and the tidal field eigenvector with the smallest eigenvalue. Random alignment corresponds to a value of $1$ in this figure. The distributions are shown for the full population of galaxies in the Horizon-AGN simulation at different redshifts.}
\label{fig:fullpop}
\end{figure}

First, we describe the alignments for the full population. These results provide us with a point of reference for when we analyse the alignment of elliptical progenitors in Section \ref{sec:mainres}. Figure \ref{fig:fullpop} shows $f(u)=1+\xi(u)$,
the binned alignment probability density distribution for the full population as defined in Section \ref{sec:methods}, at selected redshifts. Departures from unity indicate a significant alignment of the major axis of galaxies with ${\bf v}_1$. If the major axis of the population is aligned in the direction of ${\bf v}_1$, as an example, then smaller angles between them are more likely than larger ones. The pdf of $1+\xi(u)$ will then be broadly increasing with $\cos(\theta_1^{\rm major})$, as can be seen in figure \ref{fig:fullpop} for the $z = 0.06$ curve. It is anticipated that these curves will generally cross the value 1 at around  $\cos(\theta_1^{\rm major})=0.6$. This can be interpreted as either the average angle of a uniform random distribution ($\theta_1^{\rm major}=1\,{\rm rad}$), or the median angle of a uniform distribution ($\theta_1^{\rm major}=60 \deg$). Either way this crossing point is as expected. This can be seen for the low redshift curves. This trend evolves with redshift, transitioning to perpendicular alignment at high redshift. At $z = 3$ the population aligns perpendicularly to ${\bf v}_1$, with galaxies more often displaying values of $\cos(\theta_1^{\rm major})<0.6$.

This alignment transition can be explained by the morphological evolution and the mass build-up of the galaxy population in the simulation. 
In a recent work, \citet{Codis18} have shown both variables play a role in determining the relative alignment of a galaxy with respect to the nearest filament of the cosmic web. Their work identified a mass transition for alignment, by which galaxies above a certain mass threshold change the orientation of their spin to align perpendicularly to the direction of nearby filaments, in line with theoretical predictions \citep{codis15b}. At fixed stellar mass, there is a residual alignment trend on galaxy morphology, with disc galaxies aligning the direction of their spin along cosmic filaments and massive ellipticals, perpendicularly to them. We thus find an analogous phenomenon for galaxy shapes, with the minor axis playing the role of the spin (see Appendix \ref{app:ms}).

The galaxies in the simulation are largely disc-like at high redshift. Major and minor merger events \citep[e.g.][]{Welker2017, Martin2018_mergers,Kaviraj2014}, together with AGN feedback \citep{Dubois16} drive the transition to a more massive population with a higher fraction of ellipticals. At $z = 3$ the ratio of discs to ellipticals is $8.09:1$, while this evolves to $3.86:1$ at $z = 1$ and $1:1.53$ at $z = 0.06$. Hence, the results of Figure \ref{fig:fullpop} can be explained by the findings of \citet{Codis18}. The morphological evolution of the sample and the progressive build-up of stellar mass towards low redshift can thus explain the transition in the major axis alignment with respect to the eigenvectors of the tidal field. As we discuss in Section \ref{sec:discuss}, this phenomenon is connected to the merger history of galaxies, which drives spin swings in the cosmic web \citep{Welker2014}.

 \subsection{Alignments of high-mass ellipticals and their main progenitors}
\label{sec:mainres}

\begin{figure}
\centering
\includegraphics[width = 1.2\columnwidth,trim=35 0 0 25bp,clip]{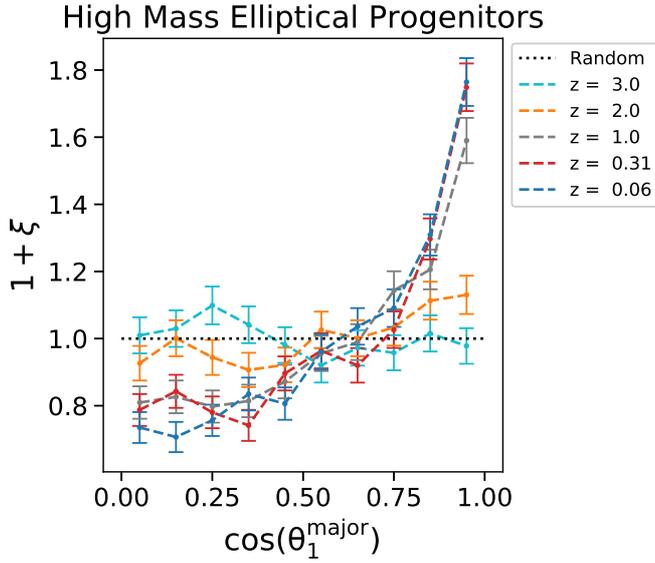}
\caption{Probability density distribution of $\cos(\theta_1^{\rm major})$, the cosine of the angle between the major axis of a galaxy and the tidal field eigenvector with the smallest eigenvalue. Random alignment corresponds to a value of $1$ in this figure. The distributions are shown for the main progenitors of the high mass $z=0.06$ ellipticals in the Horizon-AGN simulation at different redshifts.}
\label{fig:alignplots2}
\end{figure}
\begin{figure*}
	\includegraphics[width=0.495\textwidth,clip]{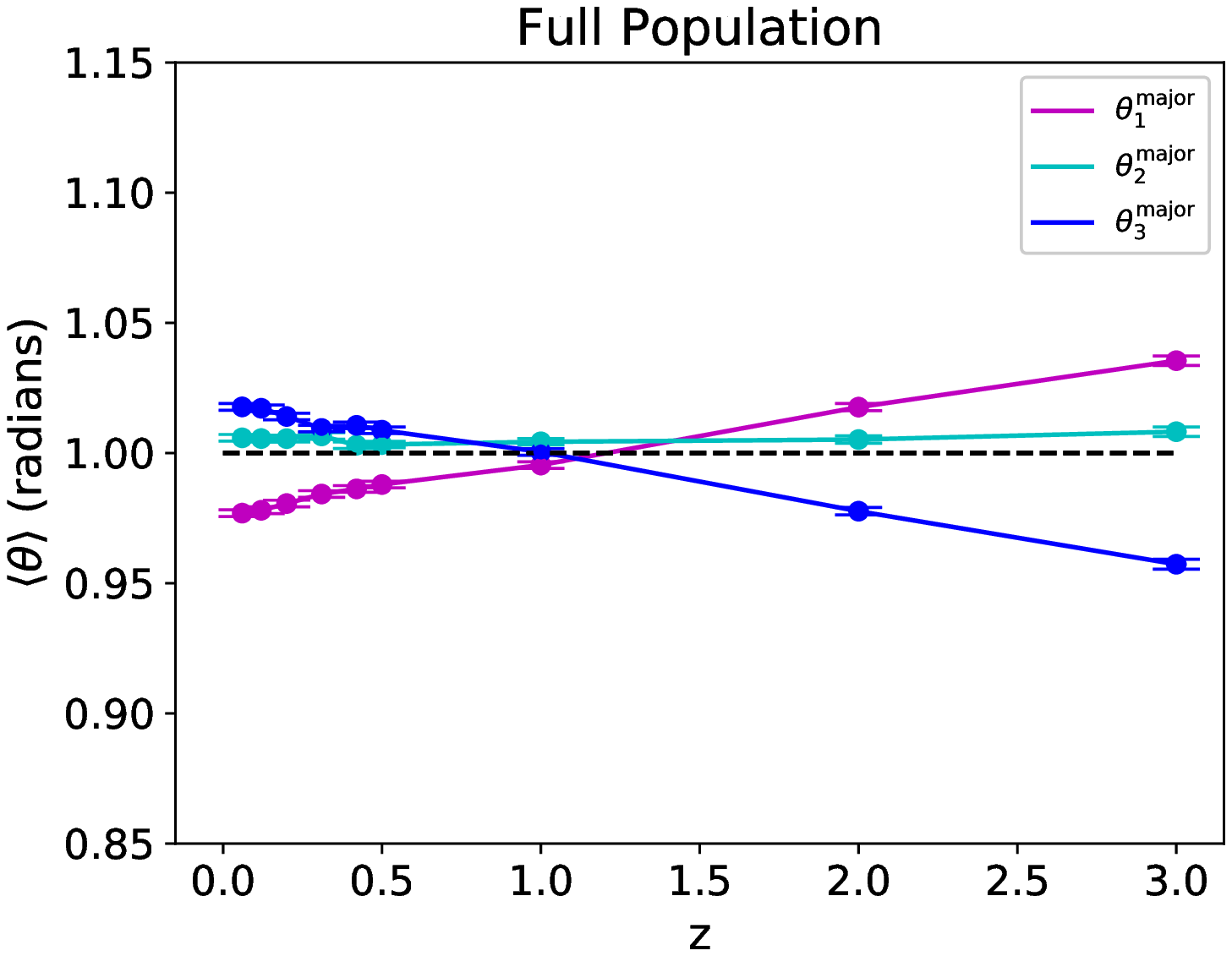}
    \includegraphics[width=0.495\textwidth,clip]{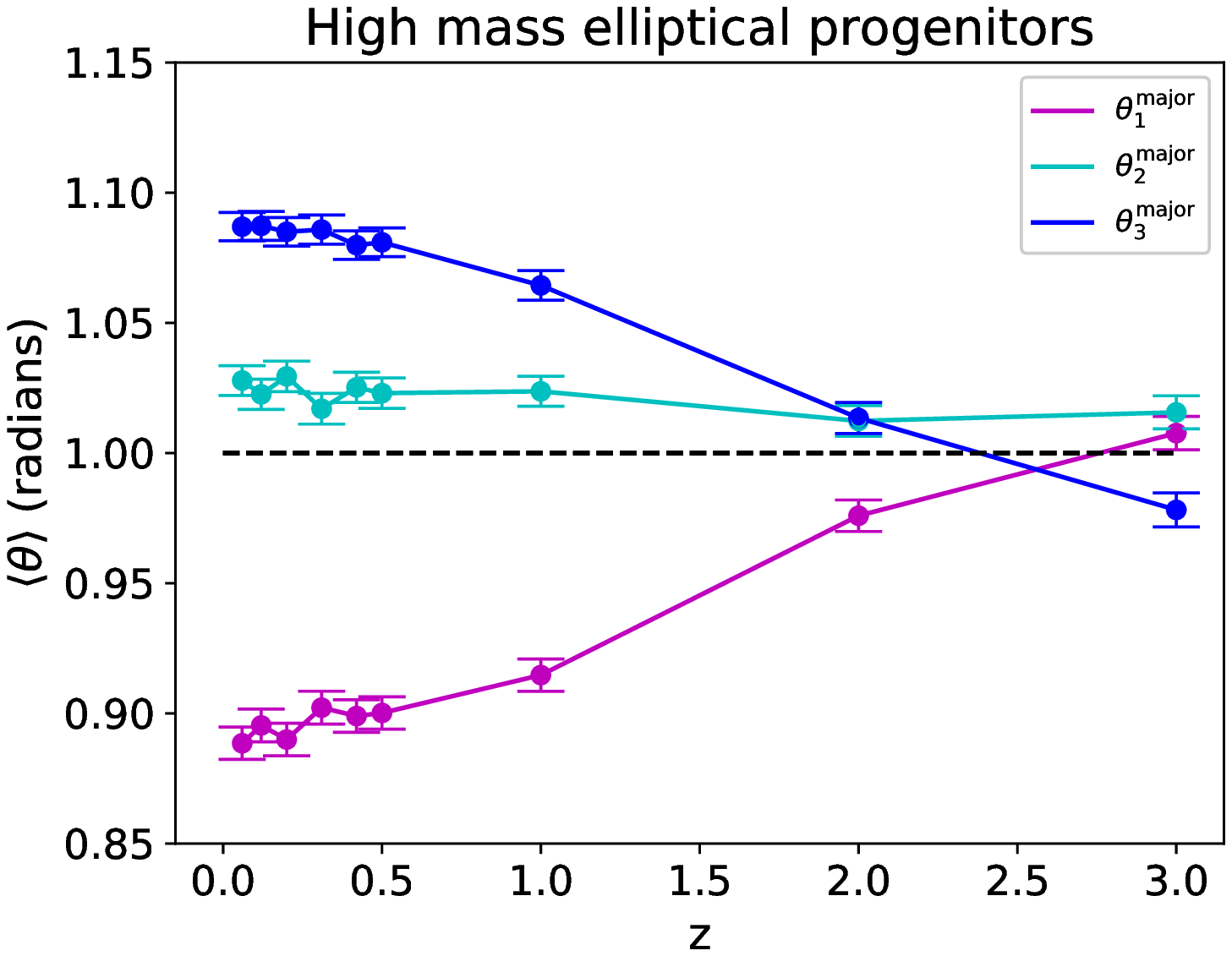}
    \caption{Average value of $\theta_1^{\rm major}$, $\theta_2^{\rm major}$ and $\theta_3^{\rm major}$ in radians as a function of redshift for: (left) the full population, and (right) the main progenitors of the high mass ellipticals. The $\langle\theta\rangle = 1$ line corresponding to a random distribution is shown for comparison (black dashed).}
    \label{fig:redshiftplots}
\end{figure*}

We now focus on the alignments with the tidal field of the sample of low redshift massive ellipticals and their main progenitors up to $z=3$. Figure \ref{fig:alignplots2} shows $1+\xi(u)$ for this population. There is a significant alignment of elliptical galaxies and their main progenitors with ${\bf v}_1$ at low redshift. An excess probability of $\sim 80\%$ is evidenced for galaxies in the highest $\cos(\theta_1^{\rm major})$ bin. This is stronger than the signal found for the full population in Figure \ref{fig:fullpop} ($\sim 15\%$), suggesting that the elliptical sub-population dominates the trend seen there at low redshift, in line with the findings of \citet{Chisari16}. 

This sample also displays an evolution in redshift of the alignment trend. However, in this case, we do not find a significant alignment at $z = 3$. This is despite the increased fraction of discs among the elliptical progenitors at this redshift, as evidenced from Figure \ref{fig:completeness}.  
To check whether this random alignment was actually a consequence of a possible cancellation of the parallel alignment of discs and the perpendicular alignment of ellipticals, we divided the sample of elliptical progenitors at $z = 3$ into elliptical and discs and measured their alignment with the tidal field separately. We thus verified that neither of those samples showed a significant alignment. Hence, the mechanism responsible for the alignment of the major axes of low redshift ellipticals with ${\bf v}_1$ must be acting between $z = 3$ and $z = 0$. We discuss this in further detail in Section \ref{sec:discuss}. Moreover, we see in Figure \ref{fig:alignplots2} that the alignment of elliptical progenitors reaches an approximate constant amplitude at $z =0.5$. 

The removal of low mass galaxies from among the $z=0.06$ elliptical sample plays a crucial role in defining the alignment trend. Without this mass cut, we have verified that the alignment evolution would be similar to that of the full population. This is consistent with the discussion presented in the previous section, whereby alignment trends are not only a function of morphology but also of stellar mass.

The probability density distribution of alignment angles shown in Figure \ref{fig:alignplots2} weights all galaxies equally. Alternatively, we considered applying the weight $1-c/a$ for each galaxy to test whether the alignment signal is dependent on ellipticity. The results were unchanged by this weighting. Galaxy ellipticity has no effect on our results.

Figure \ref{fig:redshiftplots} summarises the results of this section by showing the mean alignment angle of the massive elliptical progenitors as a function of redshift for the three different tidal field eigenvectors. The left panel shows the results for all galaxies in Horizon-AGN, while the right panel focuses on massive elliptical progenitors. Consistently with results shown in Figure \ref{fig:fullpop}, the overall population displays a transition in alignment trend at $z\sim 1$ for ${\bf v}_1$. Between $z=3$ and $z=0$, the average alignment angle of the major axes with ${\bf v}_1$ evolves from $59\deg$ to $\sim 56 \deg$. This is opposite in sign to the alignment with ${\bf v}_3$, as expected. A small positive alignment is seen with ${\bf v}_2$. High mass ellipticals (right panel) show similar trends but with a different redshift evolution. There is no significant alignment at $z=3$ but progenitors build up such alignment by $z=1$ ($\theta_1^{\rm major}\sim 51\deg$). While massive ellipticals reach a constant alignment by $z=0.5$, this is on the contrary not evidenced for the full galaxy population in Figure \ref{fig:redshiftplots}. 

\subsection{Mass-Dependence of alignments}
\label{sec:massdep}

Finally, we investigated the mass-dependence of the alignment signal of elliptical progenitors by measuring the average $\langle\theta_1^{\rm major}\rangle$, $\langle\theta_2^{\rm major}\rangle$ and $\langle\theta_3^{\rm major}\rangle$ as a function of redshift and in different bins of galaxy stellar mass. The mass bins were defined to contain an approximately equal number of elliptical galaxies at $z=0.06$, as discussed in Section \ref{sec:methods}. 

\begin{figure}
	\includegraphics[width=0.495\textwidth,clip]{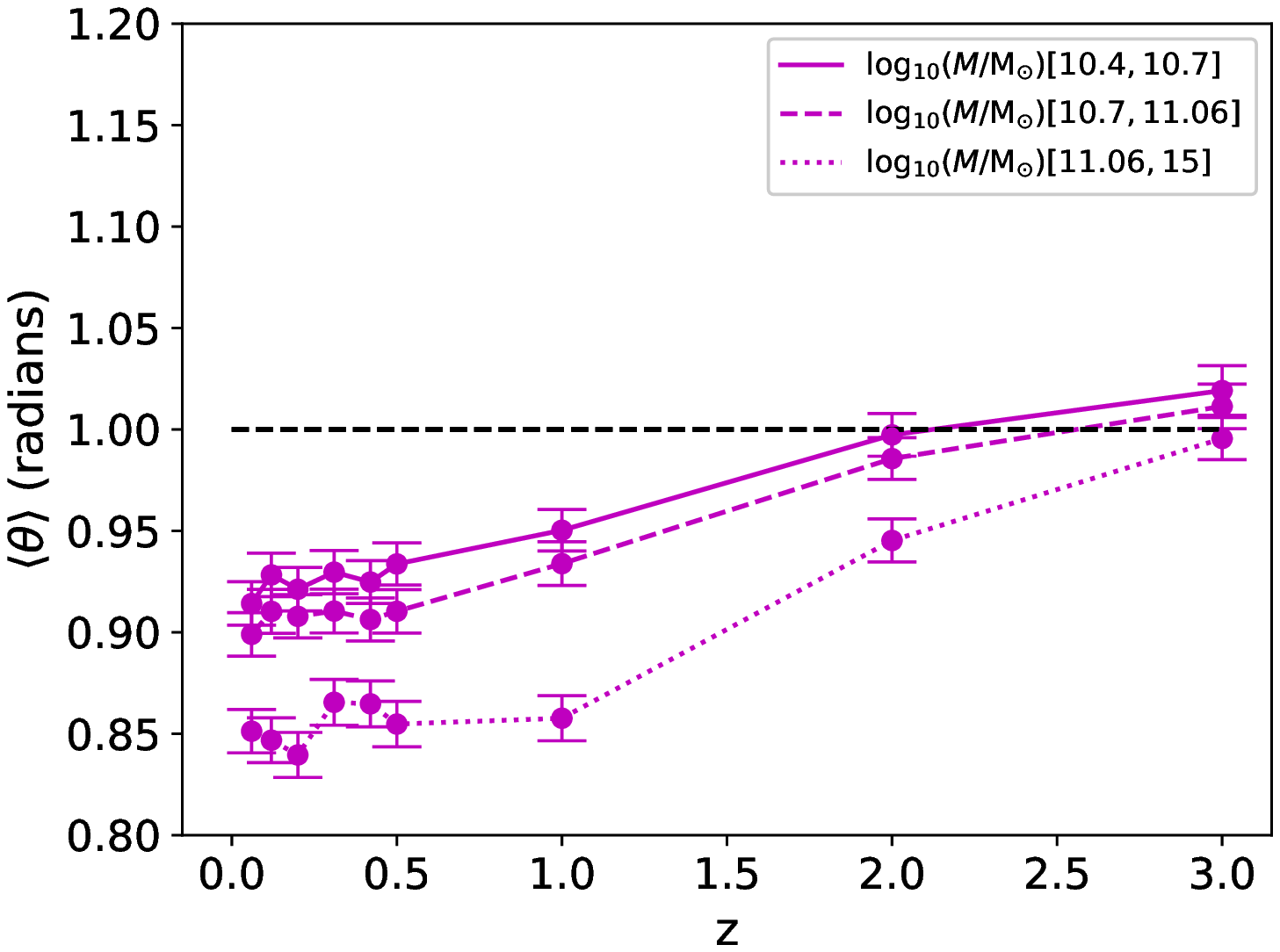}
    \includegraphics[width=0.495\textwidth,clip]{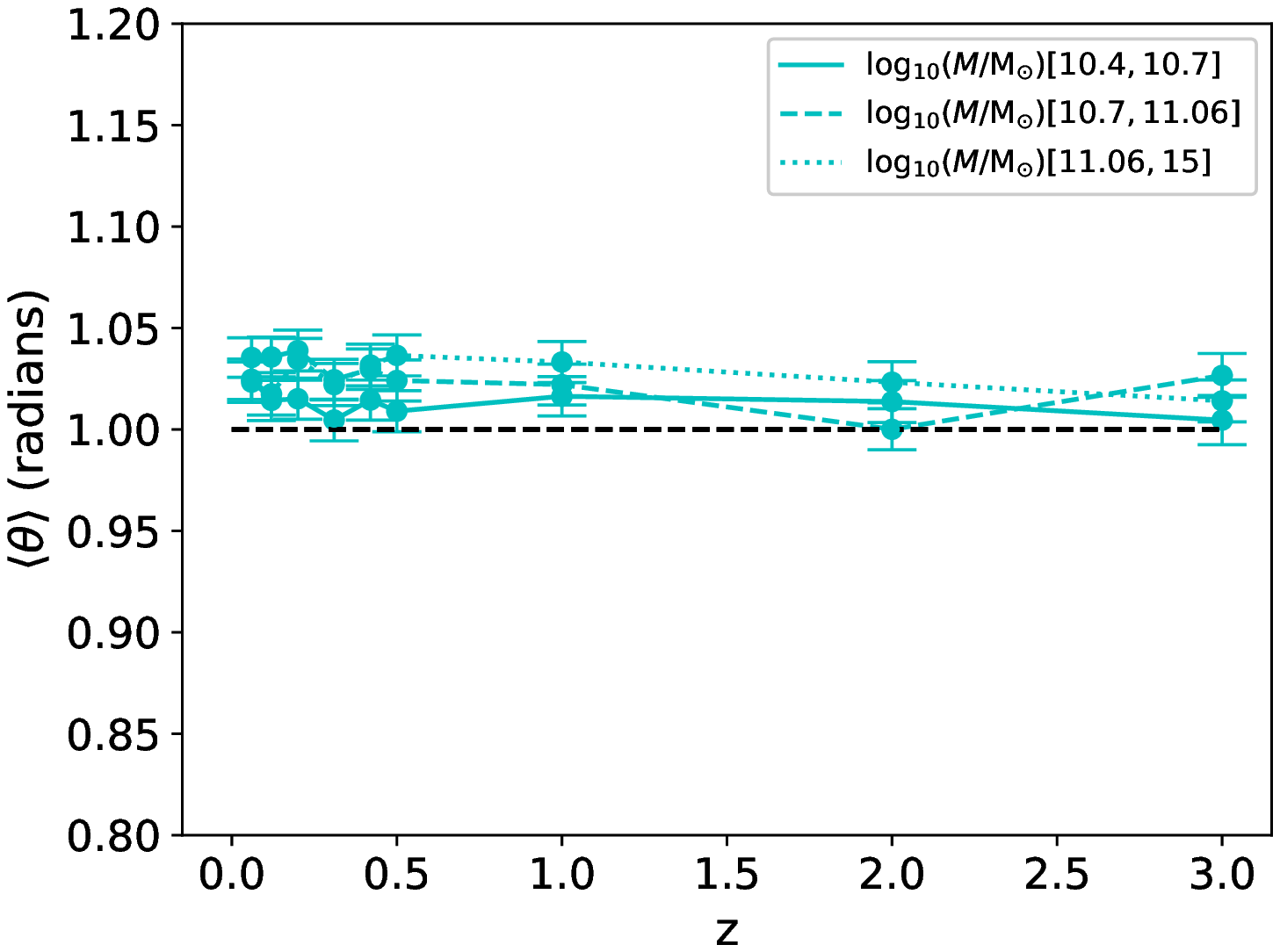}
    \includegraphics[width=0.495\textwidth,clip]{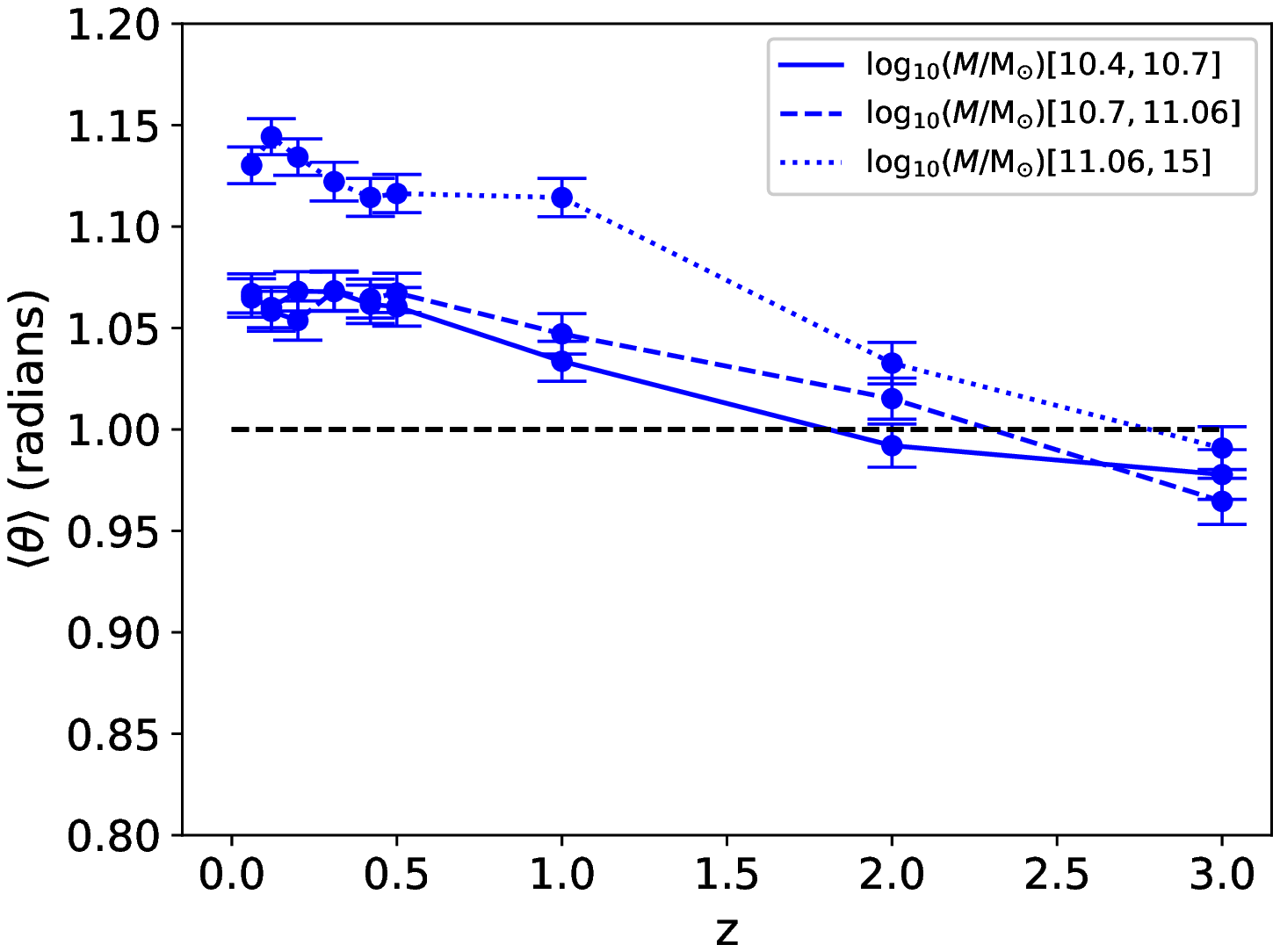}
    \caption{Average value of $\theta_1^{\rm major}$ (top), $\theta_2^{\rm major}$ (middle) and $\theta_3^{\rm major}$ (bottom) in radians as a function of redshift for galaxies in different bins of stellar mass: $10^{10.4}<M/{\rm M}_\odot<10^{10.7}$ (solid), $10^{10.7}<M/{\rm M}_\odot<10^{11.06}$ (dashed) and $M/{\rm M}_\odot>10^{11.06}$ (dotted). The value expected for random alignments is shown as the black dashed line for comparison. Alignments are stronger with the first and third eigenvectors of the tidal field, and stronger for higher mass progenitors at lower redshifts.}
    \label{fig:mcredshiftplots}
\end{figure}

The results are shown in Figure \ref{fig:mcredshiftplots}. There is a clear trend for main progenitors of higher mass galaxies to display a stronger alignment with the tidal field. This is evidenced for all tidal field eigenvectors in Figure \ref{fig:mcredshiftplots}, with the consequence that the alignment between the major axis of the progenitor and the eigenvector of the tidal field becomes significant only at a certain redshift, which depends on the mass of the low redshift elliptical being considered. The average mass values of the progenitor population in each mass bin evolve with redshift. Mass grows almost by a factor of $10$ between $z=3$ and today in all mass bins from mean values of $\log_{10}(M/{\rm M}_{\odot})=\{9.7,9.9,10.3\}$ in the highest redshift snapshot considered. These populations also decrease in their $V_\theta/\sigma$ mean values, in agreement with Figure \ref{fig:completeness}, from $V_\theta/\sigma=\{0.93,0.91,0.87\}$ at $z=3$ to $V_\theta/\sigma=\{0.43,0.37,0.27\}$ at $z=0.12$. This result will be further discussed in Section \ref{sec:discuss}. 

\section{Discussion}
\label{sec:discuss}

Numerous observational works have confirmed that massive elliptical galaxies have a tendency to align radially towards over-densities in the matter field \citep{Mandelbaum06,Hirata07,Joachimi11,Singh14,Singh15,Johnston18}. These works have been successful in constraining the amplitude of alignment, testing the tidal alignment model at low redshift and constraining the mass dependence of the alignment signal. Nevertheless, the redshift evolution of alignments remains poorly constrained, and little is known about {\it when} galaxies effectively gain their alignment. 

Several groups have recently led multiple efforts in modelling intrinsic alignments with cosmological numerical hydrodynamical simulations \citep{tenneti14a,Tenneti15a,Tenneti15b,Velliscig15,Velliscig15b,Codis15,Chisari15,Chisari16,Chisari2017,Hilbert17}, with the goal of constraining weak lensing contamination to existing and upcoming galaxy surveys. All of these works have succeeded in qualitatively reproducing the alignment trend of ellipticals at low redshift, though different behaviours have been identified for disc-like galaxies \citep{Chisari15,Tenneti15b,Kraljic19}. In this work, we focus on elliptical galaxies alone, for which there is good agreement, and answer the question of {\it when} these galaxies gained an alignment with the tidal field. To do this, we have used merger trees to trace back the main progenitors of low redshift ellipticals back in time through the Horizon-AGN simulation. 

Related work was performed by \citet{Welker2014}, who studied the relative orientation between galaxy angular momenta and the cosmic web of filaments of the large-scale structure in Horizon-AGN at $z=1.5$. In that work, the authors determined that mergers drive spin swings in the cosmic web. Low mass galaxies that have not suffered mergers throughout their history possess an angular momentum axis aligned with nearby filaments of the large-scale structure, while those with cumulatively more minor mergers tend to flip their spin perpendicularly to filaments. The latter trend has been confirmed observationally in different works, e.g. \citet{Tempel13} and \citet{Chen19}. In the absence of sustained mergers, the spin direction is dominated by the anisotropic in-fall of matter and constantly re-aligned with the filament \citep{aubertetal04,Laigle15}. \citet{Codis18} identified a mass transition for alignment, where galaxies above a certain mass threshold flip their spin to align perpendicularly to the direction of nearby filaments, in line with theoretical predictions \citep{codis15b}. Moreover, at fixed stellar mass, they found that the alignment trend depends on morphology, with discs tending to point their spins along filaments. In that work, it was also shown that the alignment of the minor axes of galaxies tends to be stronger in amplitude than the case of the spin.

In this work, we have shown that massive ellipticals that are today aligned with the tidal field did not display a significant alignment at $z=3$. \citet{Codis18} found that the stellar mass threshold for a transition from anti-alignment to alignment of the major axis of a galaxy with the nearest filament was approximately $M\simeq 10^{10.1\pm 0.3}\,{\rm M}_\odot$ and independent of redshift (though this assessment was limited by the lack of statistics in a hydrodynamical box of $100$ Mpc$/h$ on each side). Figure \ref{fig:masshist} indicates that the majority of the main progenitors of low redshift ellipticals are indeed above that threshold. It is thus likely that the reason that these progenitors do not display alignment at $z\simeq 3$ is that they are indeed transitioning between two modes of alignment at this redshift.  This is consistent with the results presented in Section \ref{sec:massdep}, whereby progenitors of more massive ellipticals display a more significant alignment at earlier redshifts. \citet{Welker2014} suggested this alignment mode resulted from the mass build-up by successive mergers, showing that simulated galaxies that had undergone more of these episodes displayed more prominent alignment with the nearest filament. Although the focus of that study was on galaxy angular momenta, we show in Appendix \ref{app:ms} that this is consistent with our findings. This interpretation is supported by the analyses of the mean mass of the progenitors as a function of redshift for each mass bin. From $z=3$ to $z=2$, progenitors transition from mean masses of $\log_{10}(M/{\rm M}_{\odot})=\{9.7,9.9,10.3\}$ to $\log_{10}(M/{\rm M}_{\odot})=\{10,10.2,10.7\}$. This is accompanied, as described in Section \ref{sec:massdep}, by a decrease in mean $V_\theta/\sigma$ in the progenitor population. 

While we have focused on massive ellipticals at low redshifts in particular, the constraints obtained for the alignment history of the full population of galaxies could help inform semi-analytic models of galaxy alignments, by providing a prediction of the redshift evolution of the alignment of galaxies with the tidal field. A more extended exploration of the parameter space of mass and galaxy dynamics is presented in Appendix \ref{app:mm}. This is a different approach to that of connecting the shape of a galaxy to the dark matter halo it inhabits \citep{joachimi13a,Velliscig15,Chisari2017} and has the advantages of not relying on halo extraction and being relatively insensitive to the smoothing scale of the tidal field (see appendix \ref{app:smooth}). In the future, such work can also help establish theoretical priors on the redshift evolution of the currently favoured intrinsic alignment models, namely, the linear alignment model \citep{Catelan01,Hirata04,Bridle07}. 

While this paper was under review, a manuscript by \citet{Bhowmick19} appeared in which a similar analysis is made on the MassiveBlack-II simulation. The main difference between our analysis and theirs is that we explicitly measure alignments with the tidal field, while \citet{Bhowmick19} measure one-point and two-point ellipticity-direction correlations of the progenitors of galaxies at $z=0.6$. They find that halo alignments with the density field (on scales of $1$ Mpc/$h$ comoving) decrease with time, while galaxy alignments with haloes increase with time (similar to the findings of \citealt{Chisari2017}). This is crucial to explain the measured evolution of the two-point statistics of galaxy alignments, which increases with time at small scales and decreases at large scales. They do not quote any transition in the sign of the alignment trend, which is what we would expect based on our results in Figure \ref{fig:redshiftplots}. However, it is likely this is a consequence of differences between specifications and sub-grid models between the two simulations \citep{Tenneti15b}.

\section{Conclusions}
\label{sec:conclusions}

Using the Horizon-AGN simulation, we identified an inversion in alignments of the full galaxy population with the ${\bf v}_1$ tidal field eigenvector from parallel alignment at $z = 3$ to perpendicular alignment at $z = 0.06$, an effect which we attribute to the morphological evolution and the mass build-up of the galaxy population. We found that high-mass ellipticals at $z = 0$ show stronger alignments than the full population at low redshift, and thus dominate the alignment signal. The main progenitors of high mass ellipticals evolve the alignment of their major axes with the tidal field from random at $z = 3$ to parallel with respect to ${\bf v}_1$ at $z = 0.06$. This leads us to conclude that the alignment mechanism for low redshift ellipticals acts between $z = 3$ and $z = 0.06$. We see an approximately constant level of alignment in this sub-population between z = 0.5 to z = 0.06, which is not measured in the full population. We also find a clear mass-dependence in the strength of the alignments, with higher mass progenitors displaying a stronger alignment signal than lower mass counterparts.

\section*{Acknowledgements}

We are thankful to Sugata Kaviraj for comments and to the anonymous referee for suggestions that helped improve and extend this work. NEC has been supported throughout this work by a Royal Astronomical Society (RAS) Research Fellowship. JB's work was supported by the Department of Physics, University of Oxford, and an undergraduate research bursary from the RAS. SC is partially supported by a research grant from Fondation MERAC. GM acknowledges support from the Science and Technology Facilities Council [ST/N504105/1]. This work has made use of the Glamdring cluster at University of Oxford, for which we gratefully acknowledge the European Research Council. 


\bibliographystyle{mnras}
\bibliography{references} 

\appendix

\section{Alignment of minor and spin axes and choice of shape measurement method}
\label{app:ms}

\begin{figure*}
\centering
\includegraphics[width =\columnwidth]{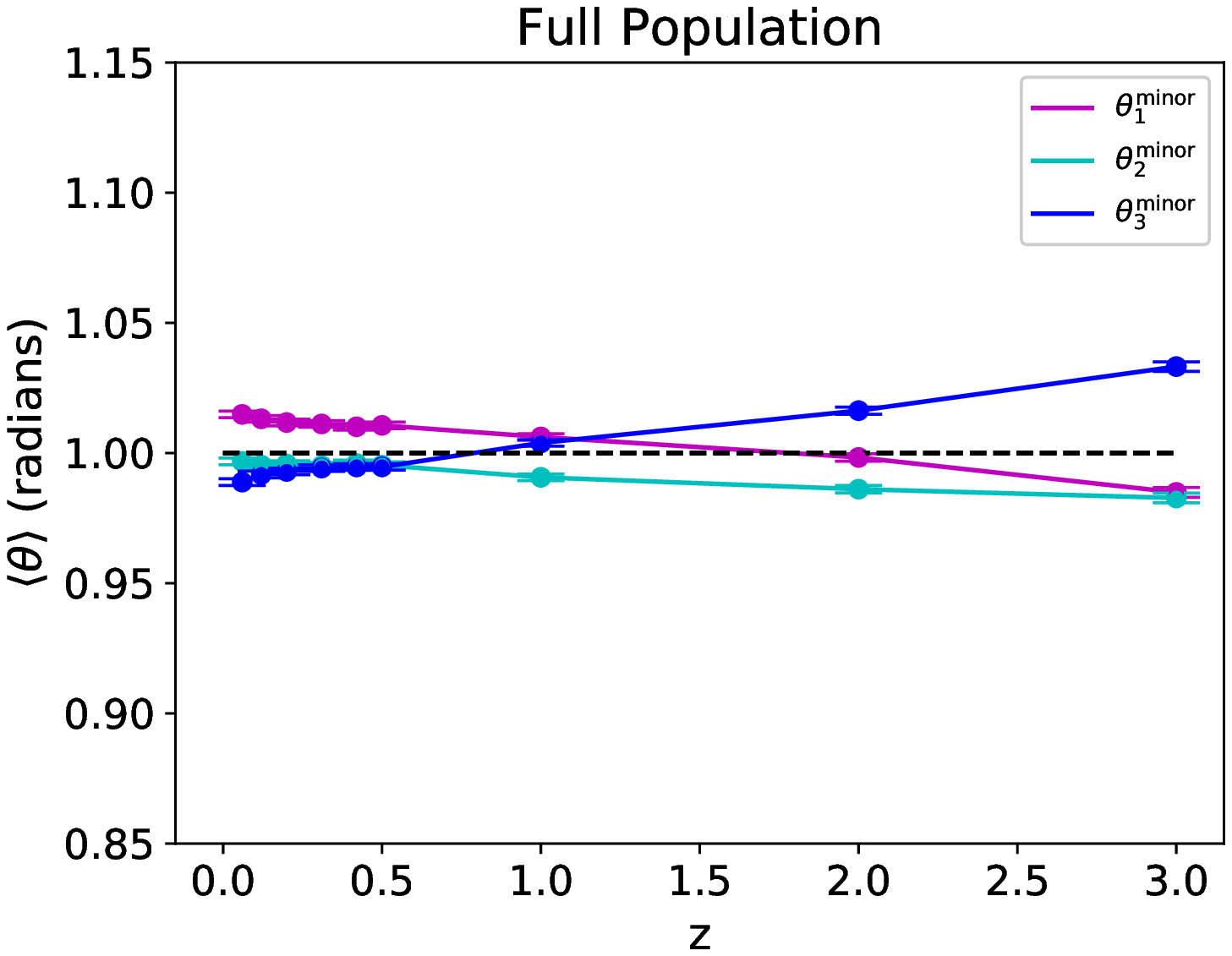}
\includegraphics[width =\columnwidth]{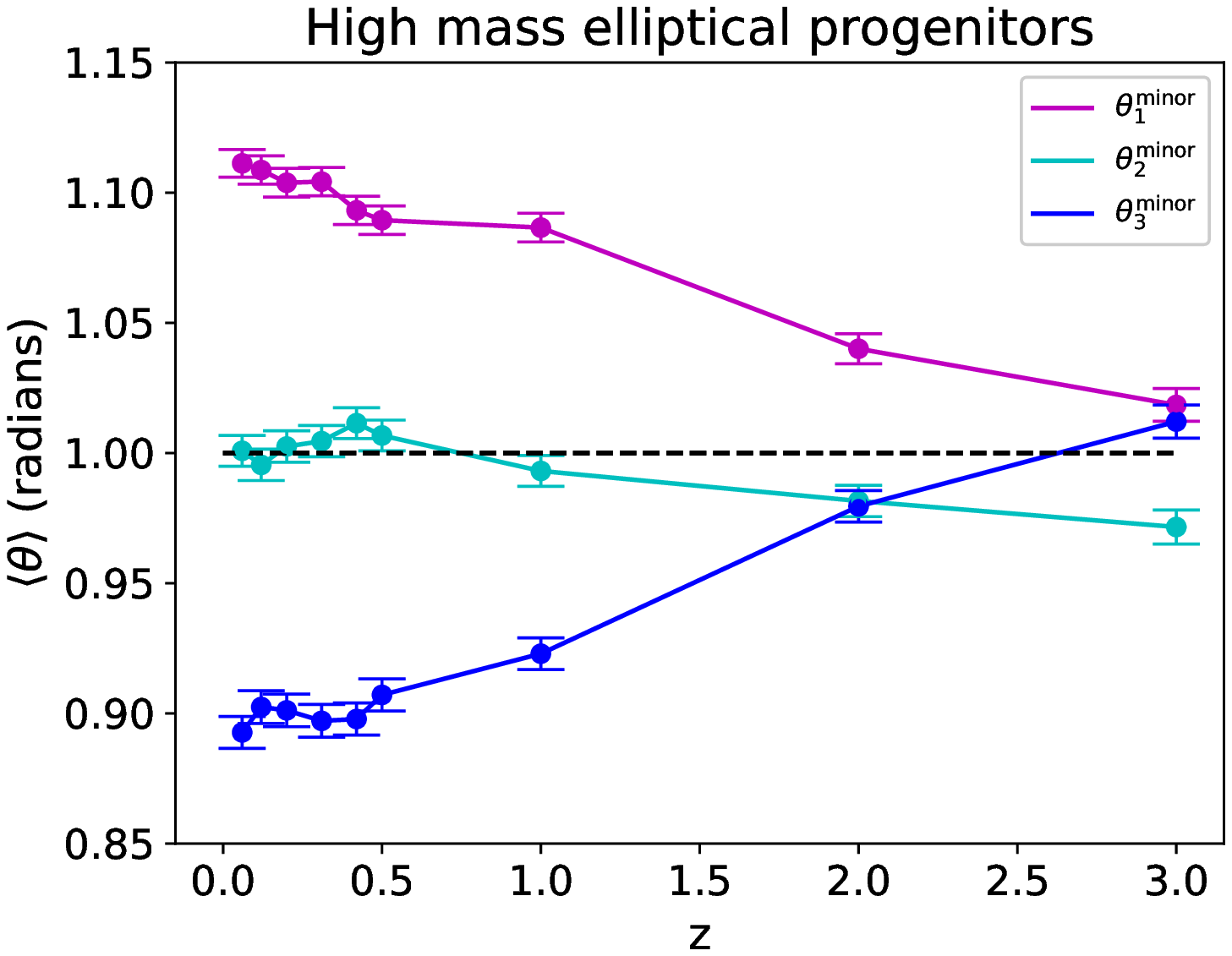}
\includegraphics[width =\columnwidth]{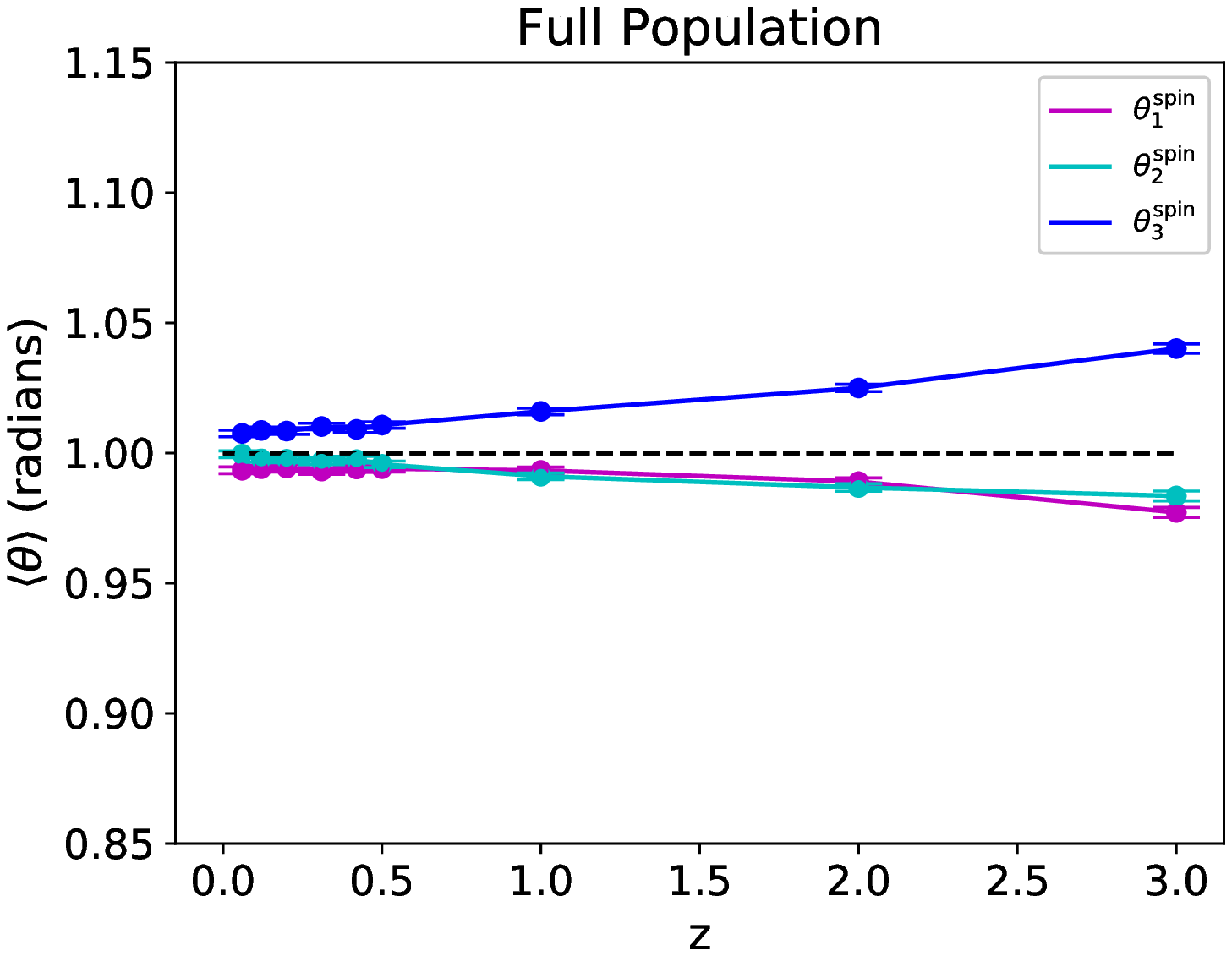}
\includegraphics[width =\columnwidth]{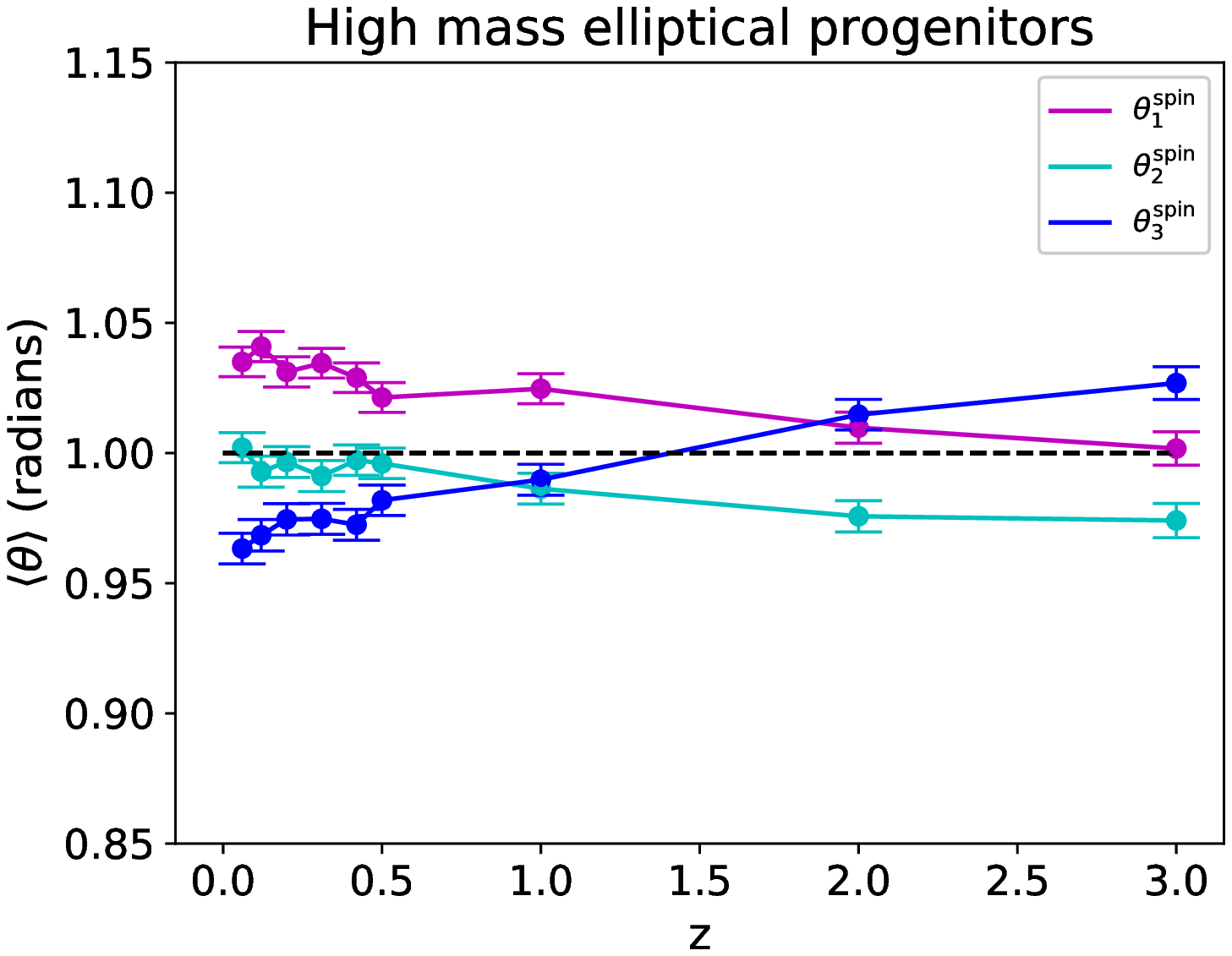}
\caption{The redshift evolution of the alignment angle for minor axes (top) and spin (bottom) of the full population of galaxies (left) and massive elliptical progenitors (right).}
\label{fig:app:ms}
\end{figure*}

We have focused on studying alignments of the major axes of galaxies in the main body of the manuscript. The reason for this choice is that in the case of ellipsoids, this direction is better defined than the minor axis, which can be degenerate with the intermediate one. For completeness, we show here results for the alignment of the minor axes, and also for the direction of the angular momenta (``spin'') of high mass elliptical progenitors in Figure \ref{fig:app:ms}.

\begin{figure*}
\centering
\includegraphics[width =2\columnwidth]{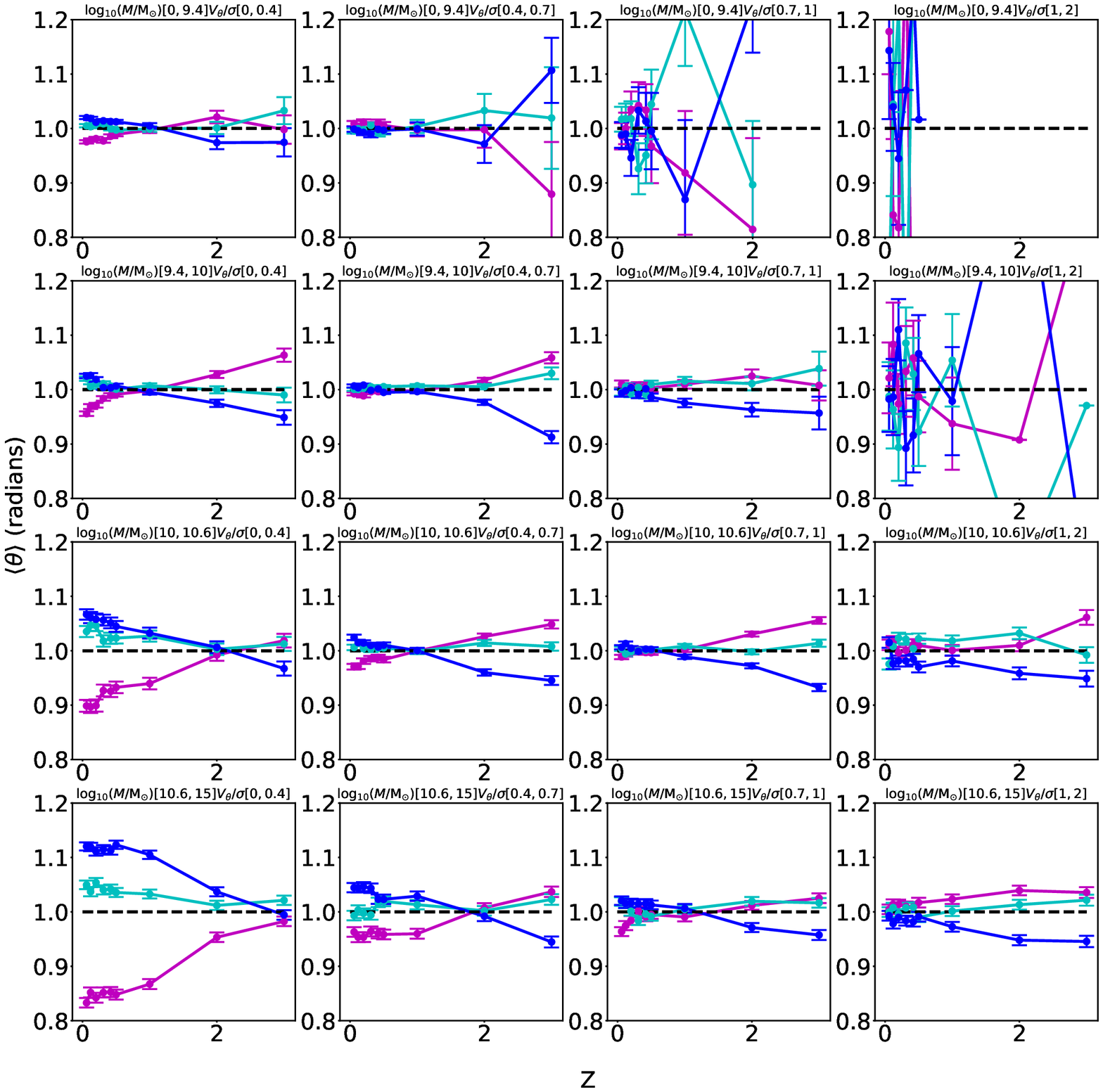}
\caption{Average value of $\theta_1^{\rm major}$ {(pink)}, $\theta_2^{\rm major}$ {(cyan)} and $\theta_3^{\rm major}$ {(blue)} in radians as a function of redshift for the progenitors of $z=0.06$ galaxies of different stellar masses and $V_\theta/\sigma$. The $\langle\theta\rangle = 1$ line corresponding to a random distribution is shown for comparison (black dashed). }
\label{fig:app:mm}
\end{figure*}

\begin{figure}
\centering
\includegraphics[width =\columnwidth]{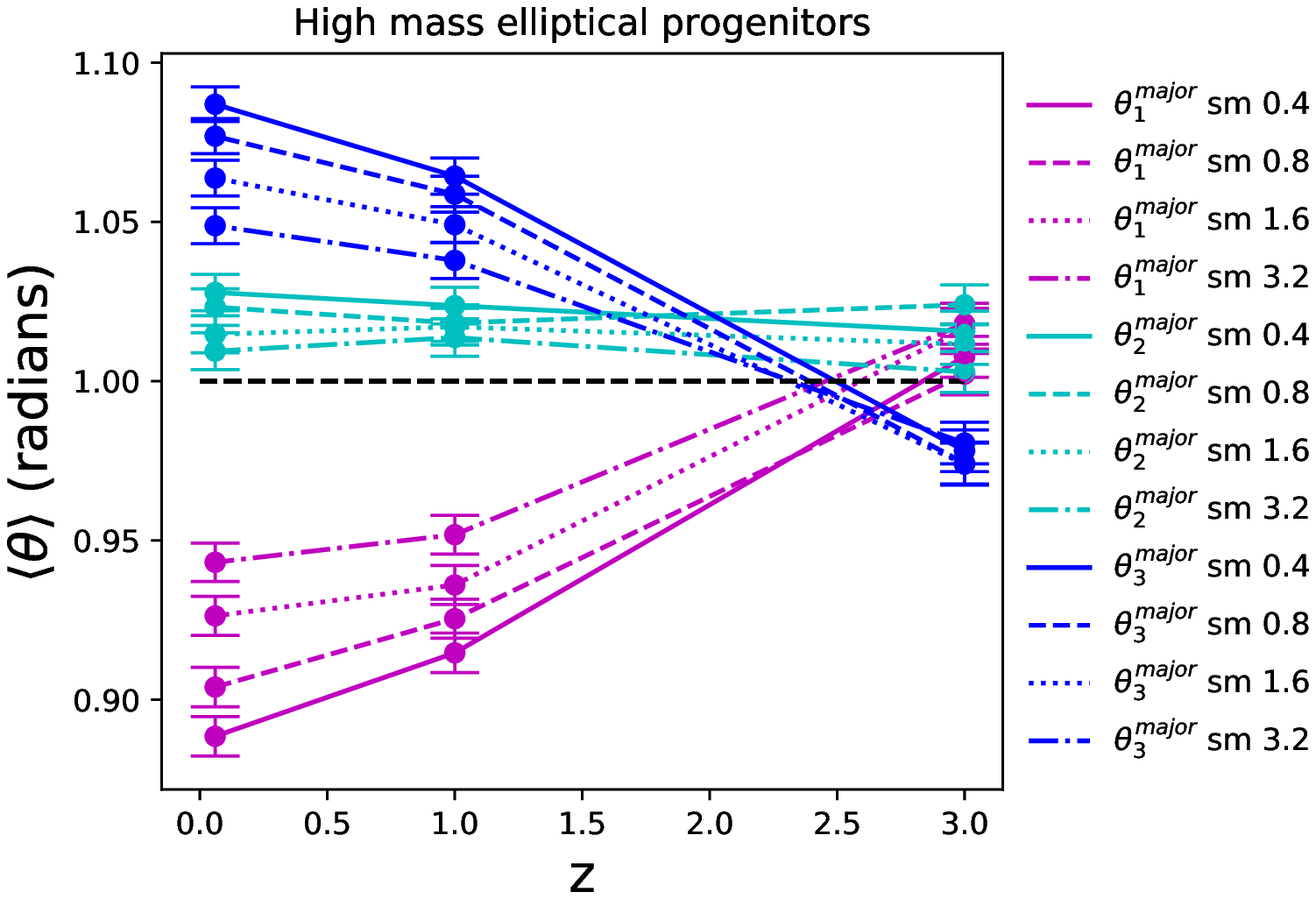}
\caption{The impact of different choice of scales for the Gaussian kernel smoothing adopted during the tidal field extraction. This figure shows the time evolution of the alignment angle of the major axes of massive elliptical progenitors with respect to the eigenvectors of the tidal field, similarly to Fig. \ref{fig:redshiftplots}. Different lines correspond to different smoothing scales: the numbers in the legend correspond to scales of $h^{-1}$ Mpc. Larger scales result in a lower alignment amplitude, in agreement with theoretical expectations and observations.}
\label{fig:app:sm}
\end{figure}

The left panels of Figure \ref{fig:app:ms} correspond to the full population, and the right panels, to massive elliptical progenitors. The top panels of Figure \ref{fig:app:ms} present results for minor axes alignments, while the bottom panels correspond to spin alignments. We see that the trends in the top panels are roughly inverted with respect to Figure \ref{fig:redshiftplots}. This is essentially a confirmation that minor and major axes are perpendicular to one another. The case of the spin is slightly different. Although the trends are qualitatively similar between spin and minor axes (top and bottom right panels), indicating a good statistical correlation between the direction of the two vectors (as demonstrated in \citealt{Chisari2017}, Figure 15), the alignment of $\theta_1^{\rm spin}$ is not as strong and it shows a later transition redshift compared to $\theta_1^{\rm minor}$. This lower strength of alignment is expected, as ellipticals have a less well-defined spin axis. This is likely a consequence of an evolution of the correlation between minor and spin axes in the progenitor population at high redshift.

To support our interpretation, we have explicitly computed the mean cosine of the angle between spin and minor axes ($\delta$), and spin and major axes ($\phi$), as a function of redshift for the progenitors of massive ellipticals. We find that spin and minor axes show a very strong correlation, with $\langle \cos\delta\rangle \simeq 0.8$. The correlation is slightly enhanced for lower mass galaxies, but present for all progenitors and throughout the full redshift range probed. These correlations does not evolve significantly below $z=1$, but shows signs of stronger alignment at $z=3$, where the population is mostly comprised of discs (Figure \ref{fig:completeness}). Opposite trends were found for the major axes, with a perpendicular alignment with the spin axes for all progenitors at all redshifts ($\langle \cos\phi\rangle \simeq 0.2$), and more strongly so at $z=3$.

The results for the alignment of the minor axes of the full population (top left panel of Figure \ref{fig:app:ms}) is also inverted with respect to Figure \ref{fig:redshiftplots}. The case of the spin is different, showing an alignment trend that is opposed to that of massive elliptical progenitors. As in the case of massive elliptical progenitors, the spin lags behind the alignment of the minor axis.

In the main body of the manuscript, we made a choice to determine minor and major axes of a galaxy using the simple inertia tensor (Eq. \ref{eq:sit}). We based this choice in that it maximized signal-to-noise of the alignments in the simulation and in the fact that the question we are addressing, that of the time evolution of alignments, is purely theoretical. We recognize, however, that shapes measured in observations tend to put more weight towards the centres of galaxies, where there is higher luminosity. There is observational \citep{Singh15,Georgiou18,Georgiou19}, as well as numerical \citep{tenneti14a}, evidence that this decreases the amplitude of alignments. Schemes that up-weight the inner regions of galaxies result in rounder shapes and lower alignment amplitudes. In general, they also result in a better correlation between the orientation of spin and minor axes. Studies of the impact of shape measurement choice in Horizon-AGN were presented in \citet{Chisari15} and \citet{Chisari16} and we refer the reader to those works for more details.

\section{Morphology- and mass-dependence of alignments}
\label{app:mm}

While the core of our work has focused on studying the progenitors of massive elliptical galaxies, we extend this in this appendix to the full parameter space of stellar mass and morphology and present analogous results to those of Figure \ref{fig:redshiftplots} to distinguish between the influence of these two galaxy properties. Figure \ref{fig:app:mm} shows the average alignment angles of the minor axes of galaxies with the tidal field eigenvectors as a function of redshift. We choose to show the minor axes alignments here due to this being better correlated with the direction of the spin of discs.

Several trends are evident from Figure \ref{fig:app:mm}. The left column shows results for ellipsoids of growing stellar mass from top to bottom. In agreement with Section \ref{sec:massdep}, the progenitors of galaxies of lower mass display a later transition in the sign of their alignment. The progenitors of galaxies at the higher end of the mass range probed by the simulation have not transitioned since $z=3$.

If we focus on the intermediate mass range (second row of panels \ref{fig:app:mm}), we see that, at fixed stellar mass, the progenitors of galaxies with higher $V_\theta/\sigma$ show trends in their alignment that are distinctive from those of ellipsoids in several ways. The $\theta_1^{\rm major}$ angle of alignment shows no transition, but it has the opposite sign to that of the progenitors of high mass ellipticals (compare to Figure \ref{fig:app:ms}). The alignment of $\theta_1^{\rm major}$ is suppressed and comparable to that of $\theta_2^{\rm major}$.

Finally, focusing on the right column of Figure \ref{fig:app:mm}, we see that high mass disc progenitors evidence trends that are similar to those measured for intermediate mass ellipticals. This suggests that there is complex interplay between stellar mass and the dynamical properties of galaxies that determines when they gain their alignment with the tidal field when the full population is considered. The redshift of transition of the alignment cannot be as clearly defined in the case of disc progenitors. The appearance of an alignment signal of $\theta_2^{\rm minor}$ and the disappearance of the $\theta_1^{\rm minor}$ alignment suggest that the environmental dependence of alignments cannot be neglected for this population.

\section{Choice of smoothing}
\label{app:smooth}

In Section \ref{sec:tidalfield}, we detailed the extraction of the tidal field in the Horizon-AGN simulation. This extraction is done relying on a specific Gaussian kernel smoothing scale. The results presented in the main body of the manuscript adopted a smoothing scale of $R_s=0.4$ $h^{-1}$ Mpc. Here, we show results for larger smoothing scales, namely: $R_s=0.8$, 1.6 and 3.2 $h^{-1}$ Mpc. Figure \ref{fig:app:sm} is the analogue of Figure \ref{fig:redshiftplots} for all four smoothing scales. The main results are qualitatively unchanged, although smoothing has an impact on the overall strength of the signal. The impact of the smoothing is not significant on the distribution of the cosine of the alignment angles, though more so on the average of the alignment angle is taken. Larger smoothing kernels result in a decrease of the alignment amplitude, as the galaxies are better correlated with the more local tidal field than with the tidal field at larger scales. This is expected from correlation studies of galaxy alignments in both observations \citep[e.g.][]{Singh14} and simulations \citep[e.g.][]{Chisari15}. For the particular case of Horizon-AGN, we had a previous study which looked at this in more detail: see \citet{Codis15}, their Figures 6 and 7 and the associated discussion.

\bsp	
\label{lastpage}
\end{document}